\documentclass[12pt]{article}
\usepackage{array}
\usepackage{amsmath}
\begin{document}
\newcommand{\be}{\begin{equation}}
\newcommand{\ee}{\end{equation}}
\newcommand{\bea}{\begin{eqnarray}}
\newcommand{\eea}{\end{eqnarray}}

\title{Constraint Algebra in Bigravity}
\author{Vladimir O. Soloviev~\footnote{mailto:Vladimir.Soloviev@ihep.ru}\\
{\small A.~A.~Logunov Institute for High Energy Physics, }\\
{\small NRC Kurchatov Institute,}\\
{\small 142281, Protvino, Moscow region, Russian Federation}}
\maketitle

\begin{abstract}
The constraint algebra is derived in the second order tetrad Ha\-mil\-to\-ni\-an formalism of the bigravity. This is done by a straightforward calculation without involving any insights, implicit functions, and Dirac brac\-kets. The tetrad approach is the only way to present the bigravity action as a linear functional of lapses and shifts, and the
Hassan-Rosen transform (characterized as ``a complicated redefinition of the shift variable'' according to the authors) appears here not as an ansatz but as  fixing of a Lagrange multiplier.  A comparison of this
approach with the other ones is provided.
\end{abstract}


\newpage
\section{Introduction}
The theory of bigravity based on de Rham, Gabadadze, Tolley (dRGT) potential~\cite{dRGT, dRGT2}  is under active studyfor about 10 years, so its general structure  is widely known, and the detailed explanation  may be found in reviews, see for instance~\cite{Reviews}. The prehistory of it will be reminded in Section 4. 
The topic of this article is the Hamiltonian approach to the bigravity and the constraint algebra of it. This subject was also discussed in many publications where various methods and variables were used. Here we are trying to combine them and show that the most transparent  picture seems to appear in the tetrad variables in the second order formalism.

The main problem of the Hamiltonian approach to massive gravity and bigravity was to prove the absence of ghosts in the theory. A scheme of such a proof by a non-perturbative method in the metric Hamiltonian formalism was proposed for the first time in a set of articles by Hassan, Rosen et al~\cite{HaRo, HaRo2, HaRo3, HaRo4}. They found a transform which provided linearity of the action in both the two lapse functions denoted here as $N,\bar{N}$ and in one shift vector $N^i$. But this transform was reasonably characterized by one of its authors as {\it a complicated redefinition of the shift variable}~\cite{HiRo}.

As it was uncovered by Dirac~\cite{Dirac1950,Dirac} and stressed by Arnowitt, Deser, and Misner (ADM)~\cite{ADM}  to make the mechanical action invariant under time reparametrizations $t \rightarrow t'=f(t)$ one should provide the Lagrangian having the 1st order homogeneity  in all its velocities, i.e. satisfying the condition $L(q_i, c\dot{q}_i)=cL(q_i, \dot{q}_i)$, where the physical time should become one of the coordinates. This requires an introduction of the lapse function  as a Lagrange multiplier
$N(t)=\dot{f}\equiv{df}/{dt}$, and the Hamiltonian acquires a form ${\rm H}=NR(q_i,p_i)$ where $R$ occurs a constraint. In the field theory  one needs to add an invariance under the spatial coordinate transformations and then additionally appears a shift vector $N^i=dx^i/dt$ accompanied by new constraints  $R_i$. The form of Hamiltonian then becomes ${\rm H}=\int d^nx(NR+N^iR_i)$.

In bigravity, we have two metric tensors with the signature $(-,+,+,+)$ defined on the same 4-dimensional manifold, and the action is a sum of the three terms:   two copies of the General Relativity action ${\rm I_f}$, ${\rm I_g}$, and a term providing the interaction ${\rm I_ {int}}=-\frac{2m^2}{\kappa}\int d^4x\ U_{\rm dRGT}$. Here both ${\rm I_f}$, ${\rm I_g}$ are linear functionals of the corresponding lapse and shift functions $N,N^i$, $\bar{N},\bar{N}^i$, whereas the potential does not have this property.  Nevertheless, this potential should be homogeneous of the 1st order under simultaneous multiplication of all functions $N,N^i,\bar{N},\bar{N}^i$ on the same constant multiplier. The Hassan-Rosen transformation~\cite{HaRo, HaRo2, HaRo3, HaRo4} in particular exploited this homogeneity in the proof of  the appearance of a new primary constraint which excluded the  superficial degree of freedom as accompanied by a secondary constraint noncommuting with it. This pair of constraints is necessary to avoid the ghost.

The next step forward was done in the work~\cite{HiRo} where the preference of the tetrad approach over the metric one 
which is coming from the linearity of dRGT potential in all the lapses and shifts was demonstrated. This advantage was elaborated further in publications~\cite{Krasnov, Alex} where the 1st order formalism was used, and so much more variables and 2nd class constraints appeared. On the base of the intermediate Dirac brackets, the authors succeeded in deriving the full constraint algebra. But this approach~\cite{Krasnov, Alex} unfortunately has not got a further application till now and no detailed comparison of its results with the metric approach is provided. In due term, the 2nd order tetrad formalism for the bigravity has been developed in the article~\cite{Kluson_tetrad}. In the first version of it the author declared his disagreement with the conclusions of the predecessors but later agreed with them.

In our previous work, we considered the Hamiltonian approach to bigravity in metric variables~\cite{SolTch1,SolTch2}, in tetrad ones~\cite{SolTetrad, SolTetrad2}, and  in the minisuperspace~\cite{SolCosmol}. The peculiarities of the different formalisms were that in the metric  
variables we did not use any explicit expression of the potential, and  worked with some equations this potential should fulfill, whereas in the tetrad variables calculations were limited to the minimal potential case $\beta_1\ne 0$. Only in the minisuperspace, all the derivations were  completed up to an explicit formula for the key Lagrange multiplier $u={\bar N}/N$. In all these cases the presence of the pair of 2nd class constraints was demonstrated. The coefficients in the constraint algebra were also derived. Now we will see that they are independent of the formalism, and their explicit meaning is transparent from this work. We are sorry to acknowledge some misprints in the published~\cite{SolCosmol} formula for the dRGT potential in tetrad variables. This work is  to complete the analysis of the problem and to compare the results with the ones obtained by other authors.

A recent publication by Kocic~\cite{Kocic} provided  essential progress in calculations dealing with the dRGT potential and its derivatives and in deriving the Hamiltonian equations by $3+1$ decomposition of the Lagrangian ones. The key point of the article~\cite{Kocic} is in defining and exploiting the geometric mean of the two metrics. Whereas it is very interesting mathematically, this is not necessary for the calculations of the bigravity potential. In a difference to this work, there were no Poisson brackets defined and constraint algebra calculated in~\cite{Kocic}. This algebra has been recently considered  by Hassan and Lundkvist~\cite{H-L}  in the metric formalism.
 The constraint algebra itself was known before, see articles~\cite{Krasnov, Alex, SolTch2, SolTetrad}, but in the metric formalism, the calculations were provided with the general implicit formula of the potential~\cite{SolTch2}, and in the 2nd order tetrad approach it was done only for the minimal potential case~\cite{SolTetrad} ($\beta_1\ne 0$). The algebra obtained in the work~\cite{H-L} coincides with the results of works~\cite{SolTch2,SolTetrad}. Another motivation of the article~\cite{H-L} was in extracting from the algebra of the 1st class constraints a new spatial metric for the bigravity. But this problem has not been solved there because it is found that both two induced metrics can play this role. 

The most technically difficult problem of
 all the mentioned approaches was a derivation of the secondary constraint denoted here as $\Omega$. In massive gravity, it was done for the first time by Hassan and Rosen~\cite{HaRo4}, see Eq. (3.32) there.

In the bigravity, the calculation of $\Omega$ has been completed in the articles (given in the chronological order):
\begin{itemize}
\item Alexandrov, Krasnov, and Speziale~\cite{Krasnov}, Eqs. (42)-(44),  without any derivatives, as the 1st order formalism is used;
\item Soloviev and Tchichikina~\cite{SolTch2}, Eq. (55);
\item Alexandrov~\cite{Alex}, Eq. (3.14),  without any derivatives, as the 1st order formalism is used;
\item Soloviev~\cite{SolTetrad}, Eqs. (30), (39)-(41); see also~\cite{SolTetrad2}, Eqs. (38), (47)-(49);
\item Hassan and Lundkvist~\cite{H-L}, Eqs. (3.18), (3.22);
\item Kocic~\cite{Kocic}, Eq. (3.8);
\item Soloviev (this work): Eq. (\ref{eq:Omega}).
\end{itemize}

This work is organized as follows. In Section 2 we present a brief introduction to the Kucha\u{r} approach in General Relativity.
Section 3 contains the basics of the Kucha\u{r} formalism for the bigravity, this formalism enjoys the explicit covariance under spacetime diffeomorphisms. In Section 4, we remind the prehistory of the bigravity theory and the construction of the dRGT potential. In Section 5, we compare the approach of this work with the pioneer method by Hassan and Rosen. Section 6 contains  a definition of the tetrad variables, the symmetry conditions, and their consequence: the Hassan-Rosen transform. In Section 7, the canonical variables are defined, and the Hamiltonian and  primary constraints are provided. Section 8 is devoted to a study of the conditions for compatibility of the primary constraints with dynamics, i.e. to a derivation of the secondary constraints and to fixing of some Lagrange multipliers. In Section 9 we formulate our conclusions. Appendix A contains expressions for the dRGT potential and the  secondary constraint $\Omega$. In  Appendix B we compare our notations with notations of related works.

We prefer to use the same notations as in
Refs.~\cite{SolTetrad, SolTetrad2, SolCosmol}. In particular, for spacetime coordinate indices running from 0 to 3, we
use small Greek letters; for internal indices running from 1 to 3, we
use small Latin letters from the beginning of the alphabet, for spatial indices small letters from the middle of the alphabet are used, for internal indices running from 0 to 3 the capital Latin letters are used. Of course, we consider only such couples of metrics that have common timelike and spacelike vectors. When the same letter is used for
analogous quantities constructed with the first ($f_{\mu\nu}$) or with
the second ($g_{\mu\nu}$) metric, then an upper bar refers to the second
one. Some additional notations are explained in Appendix A.

\section{Kucha\u{r}'s formalism in General Relativity}
The General Relativity (GR) was transformed into the Hamiltonian form in articles by Dirac~\cite{Dirac1958,Dirac1959-1,Dirac1959-2} and Arnowitt, Deser, Misner~(ADM)~\cite{ADM1959,ADM1960}. As the most suitable geometrical choice of variables has been proposed by the last authors the formalism is now named after ADM. The ADM approach is based on a special coordinate system chosen for spacetime.
 The time coordinate $x^0\equiv \tau$ is taken as a parameter for enumeration of spacelike hypersurfaces, whereas spatial coordinates $x^i$ coincide with the internal coordinate system given on hypersurfaces. The corresponding coordinate lines  $x^i=x^i(\tau)$ are worldlines of  observers that are considered at rest. Therefore if the ADM coordinates are chosen then a foliation of the spacetime  is already specified.
The ADM variables have a direct geometrical meaning,
in particular, time components of metric $f_{0\mu},f^{0\mu}$, 
and  Christoffel symbols $\Gamma^0_{ik}$ are
replaced by new variables\footnote{In order to make the notations compatible with the following sections we denote here the GR metric as $f_{\mu\nu}$.}
\be
\eta_{ij}=f_{ij},\quad N=
\frac{1}{\sqrt{-f^{00}}},
\quad N^i=\eta^{ij}f_{0j}, \quad K_{ij}=-\frac{\Gamma^0_{ij}}{\sqrt{-f^{00}}}.
\ee
 The geometrodynamics (a word coined by Wheeler, as also words lapse and shift for $N$, $N^i$) describes a motion of  spatial hypersurface through spacetime. The internal geometry of this hypersurface is given by  3-metric  $\eta_{ij}$ induced by the spacetime metric $f_{\mu\nu}$. Components of the induced metric are Lagrange coordinates. The  momenta conjugate to $\eta_{ij}$ are tensor densities proportional to the external curvature tensor
\be
\Pi^{ij}=\frac{\sqrt{\eta}}{\kappa^{(f)}}(\eta^{ij}\eta^{mn}-\eta^{im}\eta^{jn})K_{mn},
\ee
where $\kappa^{(f)}$ is a doubled coupling constant for  the interaction of metric $f_{\mu\nu}$ with matter.

Anyone who wants to reproduce a path from the  GR  Lagrangian to the Hamiltonian  given in articles~\cite{ADM1959,ADM1960} will make sure that it is not rather easy. A more transparent way to the same result was proposed by Kucha\u{r}~\cite{Kuchar1973,Kuchar,Kuchar2,Kuchar3,Kuchar4}. It is based on using  arbitrary coordinates $X^\alpha$ for the spacetime geometry given by the   metric $f_{\mu\nu}(X^\alpha)$ and on embedding functions  $e^\alpha(\tau,x^i)$ specifying a one-parametric family of spacelike hypersurfaces
\be
X^\alpha=e^\alpha(\tau,x^i).
\ee
For any hypersurface specified by a value of time coordinate $\tau=\mathrm{const}$ one gets three vector (and also covector) fields 
\be
e^\alpha_i=\frac{\partial e^\alpha}{\partial x^i},\quad e_{\alpha i}=g_{\alpha\beta}e^\beta_i,
\ee
which can be used as a basis for spacetime vectors tangential to the hypersurface. Then one may introduce a unit normal to the hypersurface specified by the following equations
\be
n_\alpha e^\alpha_i=0,\quad n^\alpha=f^{\alpha\beta}n_\beta,\quad n^\alpha n_\alpha=-1.
\ee
Now any spacetime vector or tensor field given on the hypersurface can be decomposed by means of basis $(n^\alpha,e^\alpha_i)$ or $(n_\alpha,e_{\alpha i})$ as follows
\be
A^\alpha=A^\perp n^\alpha+A^i e^\alpha_i,\qquad 
A_\alpha= A_\perp n_{\alpha}+A^i e_{\alpha i},
\ee
where
\be
A^\perp=-A^\alpha n_{\alpha}=A_\perp,\quad A_i=A^\alpha e_{\alpha i}=A_\alpha e^\alpha_i,
\ee
and also,
\be
C^\alpha_{\  \beta}= C^\perp_{\  \perp}n^\alpha n_\beta+C^{\perp i}n^\alpha e_{\beta i}+C^i_{\  \perp}e^\alpha_i n_\beta+ C^{ij}e_{\alpha i}e_{\beta j},
\ee
where
\be
C^\perp_{\ \perp}=C^\alpha_{\ \beta} n_\alpha n^\beta,\ C^{\perp i}=-C^\alpha_{\  \beta}n_\alpha e^{\beta i},\  C^i_{\  \perp}=-C^\alpha_{\  \beta}e_\alpha^i n^\beta, \  C^{ij}=C^\alpha_{\ \beta} e_{\alpha}^ie^{\beta j}.
\ee 
The Latin indices are moved down and up by the induced metric 
\be
\eta_{ij}=f_{\alpha\beta}e^\alpha_i e^\beta_j,
\ee
and its inverse $\eta^{ij}$.
The external curvature tensor is defined as follows
\be
K_{ij}=-e^\alpha_i n_{\alpha;\beta}e^\beta_j,
\ee
where the semicolon denotes a covariant derivative defined in spacetime and compatible with metric $f_{\mu\nu}$.
The lapse and shift functions are determined as follows
\be
N^\alpha(\tau,x^i)\equiv\dot e^\alpha(\tau,x^i)=Nn^\alpha+N^ie^\alpha_i.
\ee
There are important relations for projections of the covariant derivative called Gauss-Weingarten equations
\be
 e^\alpha_{i;\beta}e^\beta_j=-K_{ij}n^\alpha+\gamma_{ij}^ke^\alpha_k,
\ee
where 
$\gamma^k_{ij}$ are Christoffel symbols of the internal Riemannian geometry in the hypersurface. The  covariant derivative in this geometry is defined as follows
\be
A^i_{\ |j}=e_{\alpha}^i(A^k e^\alpha_k)_{;\beta} e^\beta_j.
\ee
The projections of the commutator of the second covariant derivatives provide Gauss-Peterson-Codazzi equations 
\bea
R^{(f)}_{mijk}&=& R^{(\eta)}_{mijk}+K_{ik}K_{mj}-K_{ij}K_{mk},\\
R^{(f)}_{\perp ijk}&=&K_{ij|k}-K_{ik|j},
\eea
and similar relations
\be
NR^{(f)}_{i\perp j\perp}=\delta_N K_{ij}+NK^m_i K_{mj}+N_{|ij},
\ee
where
\be
\delta_N K_{ij}=Nn^\mu (K_{ij})_{,\mu}.
\ee
All these results allow to express the GR Lagrangian in $3+1$ form
\be
\sqrt{-f}R^{(f)}=N\sqrt{\eta}(R^{(\eta)}-K^2+\mathrm{Sp} K^2)+(-2\sqrt{\eta}K)_{,0}+(2\sqrt{\eta}(KN^i-N^{|i}))_{,i}.
\ee
This form of Lagrangian provides an easy way to determine momenta and to make the Legendre transformation giving the Hamiltonian.
In fact, one as usual  follows this way~\cite{York,Sol1988,Gourg,Kocic,Kocic2} now.

In bigravity we deal with two different spacetime  metrics  $g_{\mu\nu}$ and $f_{\mu\nu}$, therefore we have two induced 3-metrics $\gamma_{ij}$, $\eta_{ij}$ defined on each hypersurface,  two different unit normals $\bar n^\alpha$, $n^\alpha$, and two external curvature tensors $\bar K_{ij}$, $K_{ij}$.  Of course, the mixed spacetime tensor $X^\mu_\nu=g^{\mu\alpha}f_{\alpha\nu}$ can be written through the coordinate components but it would be much simpler to use one of the two partially orthonormal bases $(\bar n^\alpha,e^\alpha_i)$  or $(n^\alpha,e^\alpha_i)$. This  reduces the number of components from 10 to  6 for the corresponding metric,  let it be $f_{\mu\nu}$, then
\be
f_{\mu\nu}=-n_\mu n_\nu+\eta_{ij}e^\mu_i e^\nu_j.
\ee

\section{Kucha\u{r}'s formalism in bigravity}
In dealing with two metrics it is suitable to replace the ADM coordinate system where $X^0$ is not arbitrary but fixed by the given spacetime foliation. Let us instead take an arbitrary frame  to save the explicit diffeo\-mor\-phism invariance, so a foliation will be prescribed by four embedding variables $X^\mu=e^\mu(\tau,x^i)$. Then the ADM frame~\cite{ADM} will be only one of the all possible embedding variables choices 
\be
X^0=\tau, \qquad X^i=x^i.
\ee
This approach was developed by Kucha\u{r}~\cite{Kuchar, Kuchar2, Kuchar3, Kuchar4}, then analogous formalism was exploited by York~\cite{York}. The lapse and shift variables  $N$, $N^i$ here can not be expressed through  $f^{00}$, $f_{0i}$ components of the metric tensor $f_{\mu\nu}$,  now they are determined by the following equations
\be
N=-\dot e^\alpha n_\alpha,\qquad N^i=\dot e^\alpha e_\alpha^i,
\ee
where
\begin{alignat}{3}
&e^\alpha_i=\frac{\partial e^\alpha}{\partial x^i}, &\qquad &e_\alpha^i=f_{\alpha\beta}e^\beta_j \eta^{ij},&\qquad &\dot e^\alpha=\frac{\partial e^\alpha}{\partial\tau},\\
&\eta_{ij}=f_{\mu\nu}e^\mu_ie^\nu_j, &\qquad &\eta_{ij}\eta^{jk}=\delta^k_i,&&\\
&n_\alpha e^\alpha_i=0,&\qquad &f^{\alpha\beta}n_\alpha n_\beta=-1, &\qquad &n^\alpha=f^{\alpha\beta}n_\beta.
\end{alignat}

We use Kucha\u{r}'s approach in bigravity  to obtain an explicitly covariant $(3+1)$-decomposition of the matrix ${\sf Y}^\alpha_\beta=g^{\alpha\mu}f_{\mu\beta}$,  because the invariant potentials are constructed as functions of this matrix. First, we are to  choose the basis as there are two different normals to the given hypersurface for the two metric tensors. Without losing any generality we choose here the basis formed of $f_{\mu\nu}$, and apply a notation $(n^\alpha,e^\alpha_i)$ for it. Then the corresponding metric is decomposed as follows
\be
f_{\mu\beta}=-n_\mu n_\beta+\eta_{ij}e_\mu^i e_\beta^j,\label{f_1}
\ee
where
$\eta_{ij}$ is the spatial metric induced on the hypersurface, and $\eta^{ij}$ is the inverse of  it. In the full analogy, we introduce the basis $(\bar n^\alpha,\bar e^\alpha_i)$ constructed of metric $g_{\mu\nu}$, then
\begin{alignat}{3}
&g_{\mu\beta}=-\bar n_\mu \bar n_\beta+\gamma_{ij}\bar e_\mu^i \bar e_\beta^j,&\qquad &\gamma_{ij}=g_{\mu\nu}\bar e^\mu_i\bar e^\nu_j,&&\\
&\bar e_\alpha^i=g_{\alpha\beta}e^\beta_j \gamma^{ij}, &\qquad &\bar e^\alpha_i=e^\alpha_i,&\qquad &\gamma^{ij}\gamma_{jk}=\delta^i_k,\\
&\bar n_\alpha e^\alpha_i=0,&\qquad &g^{\alpha\beta}\bar n_\alpha \bar n_\beta=-1, &\qquad &\bar n^\alpha=g^{\alpha\beta}\bar n_\beta.
\end{alignat}
Now we should calculate coefficients in the decomposition of matrix ${\sf Y}^\alpha_\beta$
\be
{\sf Y}^\alpha_\beta\equiv g^{\alpha\mu}f_{\mu\beta}={\sf Y}^\perp_\perp n^\alpha n_\beta+{\sf Y}^\perp_i n^\alpha e_\beta^i+{\sf Y}^k_\perp e^\alpha_k n_\beta+{\sf Y}^k_i e^\alpha_k e_\beta^i,
\ee
they are the following
\be
{\sf Y}^\perp_\perp={\sf Y}^\alpha_\beta n_\alpha n^\beta, \
{\sf Y}^\perp_i =-{\sf Y}^\alpha_\beta n_\alpha e^\beta_i, \
{\sf Y}^k_\perp =-{\sf Y}^\alpha_\beta e_\alpha^k n^\beta, \
{\sf Y}^k_i ={\sf Y}^\alpha_\beta e_\alpha^k e^\beta_i.
\ee
When decomposing $g^{\alpha\mu}$ in the basis $(n^\mu,e^\mu_i)$ we get
\be
g^{\alpha\mu}=g^{\perp\perp}n^\alpha n^\mu+ g^{\perp k}n^\alpha e^\mu_k+g^{i\perp}e^\alpha_i n^\mu+g^{ik}e^\alpha_i e^\mu_k,\label{g^-1}
\ee
and for $f_{\mu\beta}$, given Eq. (\ref{f_1}), we obtain
\be
f_{\perp\perp}=-1,\quad f_{i\perp}=0=f_{\perp k},\quad f_{ik}=\eta_{ik},
\ee
as basis $n^\mu,e^\mu_i$ is constructed of this metric.

Let us introduce new variables  $u$, $u^i$, that have at least three meanings: 1) they appear in the formulas relating two pairs of lapse and shift functions
\be
u=\frac{\bar N}{N},\qquad u^i=\frac{\bar N^i-N^i}{N},\label{eq:18}
\ee 
2) they appear in projecting tensor $g^{\mu\nu}$ onto basis $(n_\alpha,e_\alpha^i)$ 
\be
u=\frac{1}{\sqrt{-g^{\perp\perp}}}\equiv\frac{1}{\sqrt{-g^{\mu\nu}n_\mu n_\nu}},\qquad u^i=-\frac{g^{\perp i}}{g^{\perp\perp}}\equiv \frac{g^{\mu\nu}n_\mu e_\nu^i}{g^{\alpha\beta}n_\alpha n_\beta},
\ee
3) they are coefficients of transformation between the two bases $(\bar n_\alpha,\bar e_\alpha^i)$ and $(n_\alpha,e_\alpha^i)$
\be
\bar n_\mu=un_\mu,\qquad \bar e^i_\mu=e^i_\mu-u^in_\mu,\qquad \bar n^\mu=\frac{1}{u}n^\mu-\frac{u^i}{u}e^\mu_i.
\ee
These variables allow to write Eq. (\ref{g^-1}) in the following form
\be
g^{\alpha\mu}=-u^{-2}n^\alpha n^\mu+ u^{-2}u^k n^\alpha e^\mu_k+u^{-2}u^ie^\alpha_i n^\mu+(\gamma^{ik}-u^{-2}u^iu^k)e^\alpha_i e^\mu_k.
\ee
By contracting expressions from Eqs.~(\ref{f_1}), (\ref{g^-1})  we get
\be
\mathsf{Y}=\left(\begin{array}{cc} -u^{-2}[n^\mu n_\nu] & u^{-2}u^i[n^\mu e_{\nu i}] \\
u^{-2}u^j[e^\mu_j n_\nu] & \left(\gamma^{ij}-u^{-2}u^iu^j\right)[e^\mu_i e_{\nu j}] \\ \end{array}
\right), \label{eq:Ymatrix}
\ee 
or
\begin{alignat}{2}
{\sf Y}^\perp_\perp&=-u^{-2},& \qquad {\sf Y}^\perp_i&=u^{-2}u_i,\\
{\sf Y}^k_\perp&=u^{-2}u^k, &\qquad {\sf Y}^k_i&=\gamma^{kj}\eta_{ji}-u^{-2}u^ku_i.
\end{alignat}

\section{Bigravity and the  dRGT potential}
The bimetric theory seems to appear for the first time in  two articles by Rosen~\cite{NRosen, NRosen2}. Rosen's motivation was to define the energy-momentum tensor for the gravitational field. The second metric was fixed and even flat. Many years later spin-2 fields had been introduced in particle physics, and  they were already treated as dynamical ones~\cite{Salam, Salam2, Salam3, WZ}. The renewed interest in multi-dimensional Kaluza-Klein models and  the new problems of  dark energy and dark matter created bigravity~\cite{Kogan} in the form close to the present. It was proposed to take two GR Lagrangians with the minimally coupled matter fields and to organize their coupling employing a potential constructed as a scalar density constructed of the two metric tensors without any derivatives. Then the dynamical equations for both metrics are of the GR form where the sources are both the matter energy-momentum tensors, and the new tensors formed algebraically of the metrics.  The last ones are obtained as variational derivatives  of the potential over the corresponding metric tensor. The energy-momentum conservation law is fulfilled separately for each source.

An obstacle to applying these theories to physics was in the appearance of a ghost degree of freedom~\cite{BD}. But soon it becomes clear that there are potentials free of this difficulty. 

The dRGT potential is formed as a linear combination of the symmetric polynomials of matrix $\mathsf{X}^\mu_\nu=\sqrt{\mathsf{Y}}^\mu_\nu\equiv\left(\sqrt{g^{-1}f}\right)^\mu_\nu$,
$$
U=\sqrt{-g}\sum_{n=0}^4\beta_ne_n(\mathsf{X})=\beta_0\sqrt{-g}+\ldots+\beta_4\sqrt{-f}, 
$$
where 
\begin{align}
e_0&=1,\nonumber\\
e_1&=\lambda_1+\lambda_2+\lambda_3+\lambda_4,\nonumber\\
e_2&=\lambda_1\lambda_2+\lambda_2\lambda_3+\lambda_3\lambda_4+\lambda_4\lambda_1+\lambda_1\lambda_3+\lambda_2\lambda_4,\nonumber\\
e_3&=\lambda_1\lambda_2\lambda_3+\lambda_2\lambda_3\lambda_4+\lambda_1\lambda_3\lambda_4+\lambda_1\lambda_2\lambda_4,\nonumber\\
e_4&=\lambda_1\lambda_2\lambda_3\lambda_4,\label{eq:e_i}
\end{align}
and $\lambda_i$ are eigenvalues of $\mathsf{X}$. These symmetric polynomials may be also expressed through traces of ${\sf X}$ and of its degrees
\begin{align}
e_0&=1,\nonumber\\
 e_1&=\mathrm{Tr}\mathsf{X},\nonumber\\
e_2&=\frac{1}{2}\left((\mathrm{Tr}\mathsf{X})^2-\mathrm{Tr}\mathsf{X}^2 \right),\nonumber\\
e_3&=\frac{1}{6}\left((\mathrm{Tr}\mathsf{X})^3-3\mathrm{Tr}\mathsf{X}\mathrm{Tr}\mathsf{X}^2+2\mathrm{Tr}\mathsf{X}^3 \right),\nonumber\\
e_4&=\frac{1}{24}\left( (\mathrm{Tr} \mathsf{X})^4 -6 (\mathrm{Tr} \mathsf{X})^2 \mathrm{Tr} \mathsf{X}^2 +3(\mathrm{Tr} \mathsf{X}^2)^2 +8 \mathrm{Tr} \mathsf{X} \mathrm{Tr} \mathsf{X}^3 -6 \mathrm{Tr} \mathsf{X}^4
\right)=\nonumber\\
&=\det\mathsf{X}=\frac{\det ||F_{\mu a}|| }{\det ||E_{\mu a}|| }\equiv\frac{ \sqrt{-f} }{\sqrt{-g}}.\label{eq:traces}
\end{align}

After $3+1$-decomposition of both metrics based on the ADM~\cite{ADM}, Kucha\u{r}~\cite{Kuchar, Kuchar2, Kuchar3, Kuchar4} and York~\cite{York} methods  the potential can be expressed in the following form
\be
U=N\tilde U(u,u^i,\eta_{ij},\gamma_{ij}).\label{eq:tildeU}
\ee
Below we follow notations of works~\cite{SolTch1, SolTch2} which are as follows
\begin{align}
 V&=\frac{\partial \tilde U}{\partial u},\label{eq:V}\\
V_i&=\frac{\partial \tilde U}{\partial u^i},\label{eq:V_i}\\
W&=\tilde U-u\frac{\partial\tilde U}{\partial u}-u^i\frac{\partial\tilde U}{\partial u^i}.\label{eq:W}
\end{align}
Unfortunately, it was possible to get the explicit form of function $\tilde U$ in metric approach only for $1+1$ spacetime dimension. But in the massive gravity case Comelli et al.~\cite{Comelli2012, Comelli2013, Comelli_MayDay} working with an implicit potential function have succeeded in demonstrating\footnote{They applied the mathematical results by Leznov and Fairlie~\cite{Leznov}. }  that if the potential $\tilde U$ is a solution of the homogeneous Monge-Ampere equation 
\begin{equation}
{\rm Det}||\frac{\partial^2\tilde U}{\partial u^a\partial u^b}||=0,\label{eq:integrability}
\end{equation}
then the Hamiltonian formalism contains two second class constraints excluding the ghost degree of freedom. In bigravity the analogous result was obtained in works~\cite{SolTch1, SolTch2},  the constraint algebra was derived there on the base of the Dirac brackets. This algebra has been confirmed later in works based on the tetrad approach~\cite{SolTetrad, H-L}. Here we come to the same algebra, but now uncovering the meaning of all its coefficients.

\section{The Hassan-Rosen transform}
The first proof for the absence of ghost in bigravity with the dRGT potential has been given by the authors of works~\cite{HaRo, HaRo2, HaRo3, HaRo4}, who proposed a special transform of  variables. But an implicit function was present in this method also. The results were based on the properties of this function to fulfill some equation and to have some symmetry. 

Let us remind the idea. In our notations the Hassan-Rosen transform is as follows
\be
u^i=v^i+uD^i_{\ j}v^j,\label{eq:HR_transform}
\ee
here the mentioned implicit matrix function is   $D^i_{\ j}$ and the new variable is $v^i$. One easily obtain the conditions for it when requiring  fulfillment of the following matrix equation 
\be
\mathsf{X}=\sqrt{\mathsf{Y}},
\ee
 it is supposed that
\be
\mathsf{X}=\left(\begin{array}{cc} (-\frac{\varepsilon}{u})[n^\mu n_\nu] & \frac{\varepsilon v^j}{u}[n^\mu e_{\nu j}] \\
\frac{\varepsilon v^i}{u} [e^\mu_i n_\nu] & \left(-\frac{\varepsilon v^iv^j}{u}+\frac{1}{\varepsilon} D^{ij}\right)[e^\mu_i e_{\nu j}] \\ \end{array}
\right), \label{eq:Xmatrix}
\ee 
where $\varepsilon=1/\sqrt{1-\eta_{ij}v^iv^j}$ and $\mathsf{Y}$ is given by Eq.~(\ref{eq:Ymatrix}). The necessary conditions are
\be
D^{ij}=D^{ji}, \qquad \gamma^{ij}=D^i_kv^kD^j_mv^m+\varepsilon^{-2}D^{ik}D_k^{\ j}.\label{eq:H-R-condition}
\ee
The indices here are moved up and down by the spatial metric  $\eta_{ij}$ and its inverse. Below we obtain an explicit form of the matrix $D^i_{\ j}$ in the tetrad formalism.

\section{The tetrads in GR and bigravity}
The description of the gravitational field in metric terms  is not the only one possible. The metric can be replaced, for example, by the matrix of the tetrad variables  $E_\mu^A$ that taken together with its inverse  $E_A^\mu$  allows to use an orthonormalized basis
required for coupling gravity to fermions. The metric and the tetrads are related by the following equations
\begin{alignat}{2}
g_{\mu\nu}&=E_\mu^AE_\nu^Bh_{AB},&\qquad &h_{AB}=\mbox{diag}(-1,1,1,1),\label{eq:tetrads2}\\
g^{\mu\nu}&=E^\mu_AE^\nu_Bh^{AB}, &\qquad &E^\mu_AE_\nu^A=\delta^\mu_\nu, \qquad E^\mu_AE_\mu^B=\delta_A^B.\label{eq:tetrads1}
\end{alignat}
The Lorentz transformations (4-rotations) of tetrads
\be
\Lambda^A_{\ B}E^B_\mu={E'}^A_\mu,\qquad E^\mu_C\Lambda^C_{\ D}={E'}^\mu_D, \qquad \Lambda_A^{\ C}h_{CD}\Lambda^D_{\ B}=h_{AB},
\ee
do not change the 
metric tensor. By using this freedom of  4-rotations we can chose one of the tetrad covectors (the timelike one) to coincide with the unit normal covector of the hypersurface, for example, 
\be
E_{0\mu}=\bar n_\mu,
\ee
then $E_{a\mu}$ will be tangential to the hypersurface. It is suitable to construct a triad ${\bf e}_i^a=E_\mu^ae^\mu_i$ related to the induced metric by the following formulas  
\begin{alignat}{2}
\gamma_{ij}&={\bf e}_i^a{\bf e}_j^b\delta_{ab},&\qquad &\delta_{ab}=\mbox{diag}(1,1,1),\label{eq:gamma_triads}\\
 \gamma^{ij}&={\bf e}^i_a{\bf e}^j_a, &\qquad &{\bf e}^i_a{\bf e}_{ib}=\delta_{ab}, \qquad {\bf e}^i_a{\bf e}_{ja}=\delta^i_j.
\end{alignat}
This is usually called as a choice of the suited tetrads~\cite{HamTetrads3}.
\be
E_\mu^0=-\bar n_\mu,\qquad E_\mu^a=\bar e_\mu^i{\bf e}_{ia}.
\ee

One may argue in the opposite direction: let us first introduce triad representation for the induced metric and then lift these triad 3-vectors from the hypersurface to the spacetime 4-vectors as follows
\be
E^\mu_a={\bf e}^i_{a} e^\mu_i.
\ee
As the bigravity potential expressed in tetrads is invariant only under the diagonal rotations of the two tetrads $E^mu_A$, $F^\mu_A$ we can not take the second tetrad  $F$ a suited one also. Instead, we parameterize\footnote{Below we will see that $\varepsilon$ introduced here coincide with that introduced in Eq.~(\ref{eq:Xmatrix}).} it as a product of an arbitrary Lorentz boost 
  \begin{equation}
\Lambda^A_{ \ B}=\left(\begin{array}{cc} \varepsilon & p_b \\
p^a &{\cal P}^a_b \\ \end{array}
\right), \quad \varepsilon=\sqrt{1+p_ap^a},\quad {\cal P}^a_b =\delta^a_{ \ b}+\frac{1}{\varepsilon+1}p^a p_b\label{eq:Lorentz} \ ,
\end{equation}
on a suited tetrad ${\cal F}$
\begin{align}
{\cal F}_\mu^0&=-n_\mu,\\
{\cal F}_\mu^a&=e_\mu^j{\bf f}_{ja}, \qquad e_\mu^j=f_{\mu\nu}e^\nu_i\eta^{ij},\\
F^A_\nu&=\Lambda^A_{\ B} {\cal F}^B_\nu.
\end{align}
 The parameter of this boost, $p_a$ or $v_a=p_a/\varepsilon$, therefore will be a dynamical variable of the bigravity. We express 
 all the four vectors of the second tetrad $(F^0_i,F^a_i)$ through the three spatial vectors of a triad (formed of the suited tetrad ${\cal F}$) and of this parameter. As a result, in the new notations
\begin{align}
 F^a_i&=\tilde f_{ai}\equiv {\cal P}_{ab}{\bf f}_{bi},\\
F^0_i&=\tilde{v}_i\equiv v_a\tilde{f}_{ai}\equiv p_a{\bf f}_{ai}\equiv p_i,
\end{align}
and the corresponding induced metric is as follows
\be
\eta_{ij}=F^A_iF_{Aj}= -\tilde{v}_i\tilde{v}_j+\tilde{f}_{ai}\tilde{f}_{aj}\equiv -p_ip_j+{\tilde f}_{ai}{\tilde f}_{aj} .\label{eq:tetradF}
\ee
The Poisson brackets are canonical
\begin{alignat}{2}
\{F^0_i,\Pi^j_b\}&=0,&\qquad \{F^0_i,\Pi^j_0\}&=\delta_i^j\delta(x,y),\label{eq:tetrbra1}\\
\{F^a_i,\Pi^j_0\}&=0,&\qquad \{F^a_i,\Pi^j_b\}&=\delta^j_i\delta_{ab}\delta(x,y). \label{eq:tetrbra2}
\end{alignat}
In the practical calculations sometimes it is suitable to use noncanonical variables $v_a=p_a/\varepsilon$, ${\bf f}_{ai}$ instead of the canonical ones $F^0_i=\tilde{v}_i\equiv p_i$, $F^a_i={\tilde f}_{ai}$, and the following relations derived from Eqs.~(\ref{eq:tetrbra1}), (\ref{eq:tetrbra2}) 
\begin{align}
\{v_a(x),\Pi^i_0(y)\}&=\tilde f^{ia}\delta(x,y), \qquad \{v_a(x),\Pi^i_b(y)\}=-\tilde f^{ia}v_b\delta(x,y),\\
\{{\bf f}_{ai}(x),\Pi^j_b(y)\}&=\left(\delta_i^j\delta_{ab}+\frac{p_ip_b{\bf f}^{ja}}{\varepsilon+1} \right)\delta(x,y),\\
 \{{\bf f}_{ai}(x),\Pi^j_0(y)\}&=-\frac{p_a\delta^j_i+\varepsilon{\bf f}^{ja}p_i}{\varepsilon+1}\delta(x,y),
\end{align}
where a new notation
\be
\tilde f^{ia}=({\cal P})^{-1}_{ab}{\bf f}^{ib},
\ee
 is introduced.

It was remarked in the work~\cite{HiRo} that after replacing metric variables by tetrads it is easy to obtain an explicit expression of the dRGT potential and it is linear in all lapses and shifts of the two metrics\footnote{The celebrated transform  Eq.~(\ref{eq:HR_transform}) permits to exclude one of the shifts.}.
In fact, matrix
\be
\mathsf{X}^\mu_\nu=E^{\mu A}F_{\nu A},\label{eq:matrixX}
\ee
occurs a square root of matrix $\mathsf{Y}^\mu_\nu=g^{\mu\alpha}f_{\alpha\nu}$, if the symmetry conditions
\be
E^\mu_AF_\mu^B-E^{\mu B}F_{\mu A}=0. \label{eq:symmetry}
\ee
are fulfilled.

With the given formulas for tetrads, we can calculate matrix $\mathsf{X}^\mu_\nu$ defined by Eq.~(\ref{eq:matrixX}) and obtain the following
\begin{equation}
\mathsf{X}^\mu_\nu=\left(\begin{array}{cc} A[n^\mu n_\nu] & B^j[n^\mu e_{\nu j}] \\
C^i [e^\mu_i n_\nu] & D^{ij}[e^\mu_i e_{\nu j}] \\ \end{array}
\right), 
\end{equation}
where
\begin{align}
A&=-\frac{\varepsilon}{u},\\
B^j&=\frac{p_a{\bf f}^{ja}}{u}\equiv \frac{\varepsilon v^j}{u},\\
C^i&=\frac{\varepsilon u^i}{u}-p_a{\bf e}^{ia}\equiv \frac{\varepsilon (u^i-u\bar{v}^{i})}{u} ,\\
D^{ij}&=-\frac{u^ip^a{\bf f}^{ja}}{u}+{\bf f}^{ja}{\cal P}_{ab}{\bf e}^{ib}\equiv -\frac{\varepsilon u^i{v}^{j}}{u}+{\bf f}^{ja}{\cal P}_{ab}{\bf e}^{ib}.\label{eq:ABCD}
\end{align}
In order to calculate the symmetric polynomials of matrix $\mathsf{X}$, at first we estimate the traces\footnote{There was a misprint in the last sign of Eq.~(\ref{eq:X^3}) in the published text of our work~\cite{SolTetrad}. } 
\begin{align}
\mathrm{Tr}\mathsf{X}&=-A+D,\\
\mathrm{Tr}\mathsf{X}^2&=A^2-2(BC)+\mathrm{Tr}D^2,\\
\mathrm{Tr}\mathsf{X}^3&=-A^3+3A(BC)-3(BDC)+\mathrm{Tr}D^3.\label{eq:X^3}
\end{align}
Given expressions for all the symmetric polynomials Eq.~(\ref{eq:traces}) we can obtain an explicit formula for the dRGT potential that occurs linear in variables
 $u,u^i$
\be
\tilde U=uV+u^iV_i+W.
\ee
The formulas for $V$, $V_i$, and $W$ are given in Appendix A. These expressions depend on canonical variables ${\bf e}_{ia}$, ${\tilde f}_{ia}$, $p_i$. 
As the potential does not contain velocities, it does not change in the course of transformation from the Lagrangian variables to the Hamiltonian ones.

At last, we should pay attention to the symmetry conditions Eq.~(\ref{eq:symmetry}). In the Hamiltonian variables, they take the following form 
\begin{align}
G_a&\equiv p_a+up_b{\bf  f}_{bj}{\bf  e}^{ja}-u^j{\cal P}_{ab}{\bf  f}_{bj}=0,\label{eq:1}\\
G_{ab}&\equiv
{\bf f}_{ci} {\cal P}_{c[a} {\bf e}^i_{b]}\equiv{\cal P}_{[ac} x_{cb]}\equiv z_{ab}-z_{ba}
=0\label{eq:2}.
\end{align}

Given (\ref{eq:2}) Eqs.~(\ref{eq:1}) can be solved for $u^i$
\be
u^i={\bf f}^{ib}\left(\frac{p_b}{\varepsilon}+up_a{\bf f}_{aj}{\bf e}^{jc}{\cal P}^{-1}_{cb} \right),\label{eq:3}
\ee
where
\be
{\cal P}^{-1}_{cb}=\delta_{cb}-\frac{p_c p_b}{\varepsilon(\varepsilon+1)}.
\ee
The functions $u^i$ are possible to express\footnote{It was shown for the first time in the work~\cite{Kluson_tetrad}.} from Eqs.~(\ref{eq:2}),  (\ref{eq:3}) as follows
\be
u^i=v^i+u{\bar v}^i,\label{eq:u^izamen}
\ee
and this allows presenting the  tetrad symmetry condition in the simplest form
\begin{alignat}{2}
u^i&=v^i+u\bar v^i,& \quad \mbox{where} \quad &\bar v^i=\frac{{\bf e}^{ia}p_a}{\varepsilon}\equiv{\bf e}^{ia}v_a ,\label{eq:u^izamena}\\
z_{ij}&=z_{ji}, &\quad \mbox{where} \quad &z_{ij}={\bf e}_{ai}\tilde f_{aj}\equiv{\bf e}_{ai}z_{ab}{\bf e}_{bj} .\label{eq:u^izamena2}
\end{alignat}

Given these results matrix $\mathsf{X}$ takes the following form
\begin{equation}
\mathsf{X}^\mu_\nu=\left(\begin{array}{cc} -\frac{\varepsilon}{u}[n^\mu n_\nu] & \frac{\varepsilon v^j}{u}[n^\mu e_{\nu j}] \\
\frac{\varepsilon v^i}{u} [e^\mu_i n_\nu] & \left(z^{ij}-\frac{\varepsilon v^iv^j}{u}\right)[e^\mu_i e_{\nu j}] \\ \end{array}
\right), 
\end{equation}
where the inverse of matrix $z_{ij}$ appears
\be
z^{ij}=\tilde f^{ia}{\bf e}^{ja}=z^{ji}.
\ee
If one compares the derived expression of matrix  $\mathsf{X}$ with Eq.~(\ref{eq:Xmatrix}) it is easy to see that they are equivalent if we take
\be
D^{ij}=\varepsilon z^{ij}\equiv \varepsilon \tilde f^{ia}{\bf e}^{ja}.\label{eq:equivalent}
\ee
Then  Eq.~(\ref{eq:H-R-condition}) is satisfied as
\be
D^i_{\ j}v^j=\bar{v}^j,\qquad D^{ik}D_k^{\ j}=\varepsilon \mathbf{e}^{ia}{\cal P}^{-1}_{ab}\mathbf{f}^{kb}\eta_{k\ell}\varepsilon{\cal P}^{-1}_{cd}\mathbf{f}^{\ell d}\mathbf{e}^{jd}=\mathbf{e}^{ia}\mathbf{e}^{ja}-\bar{v}^i\bar{v}^j.
\ee
The similar formulas were obtained in Ref.~\cite{Hassan2014}, see Eqs. (3.10), (3.11) there, but without application to the Hamiltonian formalism.

\section{The constraints}
The momenta do not depend on the potential therefore we can rewrite Hamiltonians ${\mathrm H}_g$ and ${\mathrm H}_f$ in the tetrad variables and their conjugate momenta as calculated before adding the interaction between $f_{\mu\nu}$ and $g_{\mu\nu}$. These Hamiltonians are the same as they were in the GR~\cite{HamTetrads}.

In the metric approach, the Hamiltonians for $f_{\mu\nu}$ and $g_{\mu\nu}$ are as follows
\begin{align}
\mathrm{H}_f&=\int d^3x \left(N{\cal H}+N^i{\cal H}_i\right),\\
\mathrm{H}_g&=\int d^3x \left(\bar{N}\bar{\cal H}+\bar{N}^i\bar{\cal H}_i\right),
\end{align}
where
\begin{align}
{\cal  H}&={\cal H}_M-\frac{\sqrt{\eta}}{\kappa^{(f)}} \left(R^{(\eta)}-2\Lambda^{(f)}\right)-\frac{\kappa^{(f)}}{\sqrt{\eta}}\left(\frac{\Pi^2}{2}-\mathrm{Tr}\Pi^2  \right),\label{eq:constraint01}\\
{\bar{\cal  H}}&={\bar{\cal H}}_M-\frac{\sqrt{\gamma}}{\kappa^{(g)}}\gamma \left(R^{(\gamma)}-2\Lambda^{(g)}\right)-\frac{\kappa^{(g)}}{\sqrt{\gamma}}\left(\frac{\pi^2}{2}-\mathrm{Tr}\pi^2 \right),\label{eq:constraint1}
\end{align}
and
 \begin{align}   
 {\cal  H}_i&={\cal H}_{iM}-2\Pi_{i|j}^j,\label{eq:constraint02},\\
  {\bar  {\cal  H}}_i&={\bar{\cal H}}_{iM}-2\pi_{i|j}^j.\label{eq:constraint2}
 \end{align}   
${{\cal H}}_M, {{\cal H}}_{iM}$, ${\bar{\cal H}}_M, {\bar{\cal H}}_{iM}$ are the matter contributions with the minimal interaction to the corresponding metrics 
$f_{\mu\nu}$ and $g_{\mu\nu}$, $\kappa^{(f)}$, $\kappa^{(g)}$ are coupling constants of both metrics to the corresponding matter, $R^{(\eta)}$, $R^{(\gamma)}$ -- the scalar curvatures of the two metrics $\eta_{ij}$, $\gamma_{ij}$ induced on a hypersurface, $\eta=\mathrm{det}||\eta_{ij}||$, $\gamma=\mathrm{det}||\gamma_{ij}||$, $\Pi=\eta_{ij}\Pi^{ij}$, $\pi=\gamma_{ij}\pi^{ij}$, $\mathrm{Tr}\Pi^2=\Pi^{ij}\Pi_{ij}$, $\mathrm{Tr}\pi^2=\pi^{ij}\pi_{ij}$. 

Next, we are to express the canonical variables of the metric formalism in the tetrad variables. The formulas for coordinates, i.e. the induced metrics, were already given in Eqs.~(\ref{eq:gamma_triads}), (\ref{eq:tetradF}). The momenta are expressed as follows
\begin{align}
 \Pi^{ij}&=\frac{1}{4}\left(p^i\Pi^j_0+p^j\Pi^i_0+{\cal P}_{ab}({\bf f}^{ia}\Pi^{jb}+{\bf f}^{ja}\Pi^{ia}) \right),\label{eq:Pi}\\
 \pi^{ij}&=\frac{1}{4}\left({\bf e}^{ia}\pi^{ja}+{\bf e}^{ja}\pi^{ia} \right).\label{eq:pi}
\end{align}

Then the canonical variables will be projections of the tetrads $E$, $F$   on a spacelike hypersurface $X^\mu=e^\mu(\tau, x^i)$, i.e. ${\bf e}_{ai}$, $p_i$, ${\tilde f}_{ai}$, and their conjugate momenta $\pi^i_a$, $\Pi^i_0,\Pi^i_a$.
As the number of variables increases in comparison to the metric approach, there are new constraint equations that will be  generators of tetrad rotations leaving metrics invariant
\begin{align}
 L_{AB}&={F}_{iA}\Pi^i_B-{F}_{iB}\Pi^i_A=0,\label{eq:Labf}\\
 \bar L_{ab}&={\bf e}_{ia}\pi^i_b-{\bf e}_{ib}\pi^i_a=0.\label{eq:Labg}
\end{align}
We may divide 6 constraints (\ref{eq:Labf}) in two sets:
\begin{align}
L_{ab}&=\tilde{f}_{ia}\Pi^i_b-\tilde{f}_{ib}\Pi^i_a,\label{eq:L_ab}\\ 
L_{a0}&={\tilde f}_{ia}\Pi^i_0+\tilde{v}_i\Pi^i_a.\label{eq:const2}
\end{align}
The number of constraints for ${\cal L}_g$ is less than for ${\cal L}_f$, because of using suited tetrads $E^0_i=0$ there.
Then the Poisson brackets for momenta of the metric formalism appear nonzero outside the constraints surface
\begin{align}
 \{\Pi^{ij}(x),\Pi^{k\ell}(y)\}&=\frac{1}{4}\left(\eta^{ik}{\cal M}^{j\ell}+\eta^{i\ell}{\cal M}^{jk}+\eta^{jk}
{\cal M}^{i\ell}+\eta^{j\ell}{\cal M}^{ik} \right),\label{eq:PiPi}\\
\{\pi^{ij}(x),\pi^{k\ell}(y)\}&= \frac{1}{4}\left(\gamma^{ik}\bar{\cal M}^{j\ell}+\gamma^{i\ell}\bar{\cal M}^{jk}+\gamma^{jk}\bar{\cal M}^{i\ell}+\gamma^{j\ell}\bar{\cal M}^{ik} \right).\label{eq:pipi}
\end{align}
Here
\begin{align}
 {\cal M}^{ij}&=\frac{1}{4}\left(p^i\Pi^j_0-p^j\Pi^i_0+{\cal P}_{ab}({\bf f}^{ia}\Pi^{jb}-{\bf f}^{ja}\Pi^{ib})\right), \\
\bar{\cal M}^{ij}&=\frac{1}{4}\left({\bf e}^{ia}\pi^{j}_a-{\bf e}^{ja}\pi^i_a \right).
\end{align}
Also the following relations are valid:
\begin{align}
 {\cal M}^{ij}&=\frac{1}{4}L_{AB}{F}^{jA}{F}^{iB}=0, \\
\bar{\cal M}^{ij}&=\frac{1}{4}\bar L_{ab}{\bf e}^{ja}{\bf e}^{ib}=0.
\end{align}

Given the modification of Poisson brackets (\ref{eq:PiPi}), (\ref{eq:pipi}) the constraint algebra of GR in tetrad formalism differs in a presence of algebraic constraints. 
Therefore it is suitable~\cite{Henn83} to modify ${\cal H}_i$ and $\bar{\cal H}_i$ by changing Eqs.~(\ref{eq:constraint02}), (\ref{eq:constraint2}). Below we take
\begin{align}
{\cal H}_i&={\cal H}_{iM}+\Pi^k_0p_{k,i}+\Pi^k_a\tilde{f}_{ak,i}-(\Pi^k_0p_i)_{,k}-(\Pi^k_a\tilde{f}_{ai})_{,k},\\
\bar{\cal H}_i&={\bar{\cal H}}_{iM}+\pi^k_a{\bf e}_{ak,i}-(\pi^k_a{\bf e}_{ai})_{,k}.
\end{align}
Then we get
\begin{align}
\{
{\cal H}(x),{\cal H}(y)
\}&=\eta^{ik}(x)
{\cal H}_k(x)\delta_{,k}(x,y)-\eta^{ik}(y){\cal H}_k(y)\delta_{,k}(y,x),\\
\{
{\cal H}_k(x),{\cal H}(y)
\}&={\cal H}(x)\delta_{,k}(x,y),\\
\{{\cal H}_i(x),{\cal H}_j(y)\}&={\cal H}_j(x)\delta_{,i}(x,y)-{\cal H}_i(y)\delta_{,j}(y,x),\\
\{{\cal H}_k(x),L_{AB}(y)\}&=L_{AB}(x)\delta_{,k}(x,y),
\end{align}
and also
\begin{align}
\{
\bar{\cal H}(x),\bar{\cal H}(y)
\}&=\gamma^{ik}(x)
\bar{\cal H}_k(x)\delta_{,k}(x,y)-\gamma^{ik}(y)\bar{\cal H}_k(y)\delta_{,k}(y,x),\\
\{
\bar{\cal H}_k(x),\bar{\cal H}(y)
\}&=\bar{\cal H}(x)\delta_{,k}(x,y),\\
\{\bar{\cal H}_i(x),\bar{\cal H}_j(y)\}&=\bar{\cal H}_j(x)\delta_{,i}(x,y)-\bar{\cal H}_i(y)\delta_{,j}(y,x),\\
\{\bar{\cal H}_k(x),\bar{L}_{ab}(y)\}&=\bar{L}_{ab}(x)\delta_{,k}(x,y).
\end{align}
The common Hamiltonian evidently include both constraints of the metric formalism $\bar{\cal H}$, $\bar{\cal H}_i$, ${\cal H}$, ${\cal H}_i$ and the new ones (\ref{eq:Labf}), (\ref{eq:Labg})
\begin{align}
\mathrm{H}_{g+f}&=\mathrm{H}_g+\mathrm{H}_f\nonumber\\
&= \int d^3x\left(\bar N\bar{\cal H}+\bar N^i\bar{\cal H}_i+\bar\lambda^{ab}\bar L_{ab} \right)\nonumber\\
&+  \int d^3x\left( N{\cal H}+ N^i{\cal H}_i+\lambda^{AB} L_{AB} \right).
\end{align}
 Without the potential, all the constraints are first class and all the Lagrange multipliers are arbitrary. But given the potential, the bigravity action is invariant only under diagonal rotations  of the tetrads and diagonal spacetime diffeomorphisms. In particular, as we will see below, only symmetric combinations of $\bar L_{ab}$, $L_{ab}$
\be
L^+_{ab}=\bar L_{ab}+L_{ab}=0,\label{eq:Lplus}
\ee
stay first class, whereas the antisymmetric ones
\begin{align}
L^-_{ab}&\equiv\bar L_{ab}-L_{ab}=0,\label{eq:Lminus}\\ 
L_{a0}&=0,\label{eq:L0}
\end{align}
become second class. 
Given the primary constraints (\ref{eq:Lplus}) -- (\ref{eq:L0}) independent of the potential and the form of potential given in Eqs.~(\ref{eq:tildeU}) -- (\ref{eq:W}) the complete bigravity Hamiltonian is as follows
\begin{align}
\mathrm{H}&=\mathrm{H}_{g+f}+\frac{2m^2}{\kappa}\int d^3xU=\int d^3x \Biggl[N^i(
{\cal H}_i+\bar{\cal H}_i) \nonumber\\
&+ N\left( (
{\cal H}+\frac{2m^2}{\kappa}W
)
+u
(
\bar{\cal H}+\frac{2m^2}{\kappa}V
)
+u^i
(
\bar{\cal H}_i+\frac{2m^2}{\kappa}V_i
)\right)\nonumber\\
&+\lambda^+_{ab}L^+_{ab}+ \lambda^-_{ab} L^-_{ab}+\lambda^aL_{a0}\Biggr].\label{eq:H_0}
\end{align}
This Hamiltonian depends first on 9 Lagrange multipliers $\lambda^+_{ab}$, $\lambda^-_{ab}$, $\lambda^a$, next on the canonical variables, both on 42 gravitational ones, $\tilde{f}_{ai}$, $\tilde{v}_i\equiv p_i$, ${\bf e}_{ai}$ , $\Pi^i_a$, $\Pi^i_0$, $\pi^i_a$, and on the matter coordinates and momenta, at last it is linearly dependent on 8 Lagrange multipliers  $u$, $u^i$, $N$, $N^i$. The variation over $u$, $u^i$, $N$, $N^i$ gives the following equations
\begin{align}
{\cal S}'&\equiv \bar{\cal H}+\frac{2m^2}{\kappa}V=0,\label{eq:constraint_S}\\
{\cal S}_i&\equiv \bar{\cal H}_i+\frac{2m^2}{\kappa}V_i=0,\label{eq:Si}\\
{\cal R}&\equiv{\cal R}''+u{\cal S}'+u^i{\cal S}_i=0,\label{eq:R}\\
{\cal R}_i&\equiv{\cal H}_i+\bar{\cal H}_i=0,\label{eq:Ri}
\end{align}
where
\be
{\cal R}''={\cal H}+\frac{2m^2}{\kappa}W.
\ee
It follows from Eqs.~(\ref{eq:constraint_S}), (\ref{eq:Si}), (\ref{eq:R}) that
\be
{\cal R}''=0.\label{eq:R''}
\ee
Eqs.~(\ref{eq:constraint_S}), (\ref{eq:Si}), (\ref{eq:Ri}), (\ref{eq:R''}) are constraints, and they supplement  Eqs.~(\ref{eq:Lplus})--(\ref{eq:L0}) to form the full set of 17 primary constraints. The Hamiltonian is zero on the surface of these constraints, this is a necessary condition for invariance of a theory under spacetime diffeomorphisms.

\section{The  algebra of constraints}
 According to the standard Dirac procedure   to obtain a full set of constraints and to determine  the Lagrange multipliers standing at the second class constraints it is necessary to calculate the Poisson brackets of primary constraints with the Hamiltonian. To determine where a constraint is of the first or second class  one should estimate Poisson brackets between the constraints.

Thus to satisfy equations of primary constraints (\ref{eq:Lminus}), (\ref{eq:L0}) in every moment of evolution  it is necessary to fulfill some equations, and some of these equations, fortunately, occur equivalent to the conditions of symmetry (\ref{eq:1}), (\ref{eq:2}). It is enough to demonstrate this for the case of minimal potential $\beta_1\ne 0$, $\beta_2=\beta_3=0$:
\begin{align}
\dot{L}^-_{ab}&=\{{L}^-_{ab}, {\rm H}\}\approx\{L^-_{ab},\frac{2m^2}{\kappa}\int d^3x N\tilde U\}=\frac{4m^2}{\kappa}\beta_1 Neu(z_{ba}-z_{ab}),\label{eq:sc1}\\
\dot{L}_{a0}&=\{L_{a0}, {\rm H}\}\approx\frac{2m^2}{\kappa}\beta_1Ne\left(u^i\tilde{f}_{ia}-\varepsilon(v_a+uv_je^{ja})\right).\label{eq:scondition2}
\end{align}
Eqs.~(\ref{eq:sc1}) are equivalent to the first group of the symmetry conditions (\ref{eq:1}), and therefore Eqs.~(\ref{eq:1})  are secondary constraints of the Hamiltonian formalism.  Eqs.~(\ref{eq:scondition2}) which are equivalent to Eqs.~(\ref{eq:2}) determine Lagrange multipliers  $u^i$  as functions of canonical variables, and look like the Hassan-Rosen transform  
\be
u^i=v^i+u{\bar v}^i.\label{eq:u_iv_i}
\ee
In the works~\cite{HaRo,H-L} the analog of $v^i$ has been denoted as $n^i$ and treated as a new variable replacing  $(\bar{N}^i-N^i)/N\equiv u^i$, whereas ${\bar v}^i$ corresponds to $D^i_{\ j}n^j$. 

Two sets of constraints have nonzero Poisson brackets on the constraint surface, and therefore they are second class 
\begin{align}
\{L^-_{ab}(x),G_{cd}(y)\}&=\left[\delta_{ac}z_{(bd)}-\delta_{ad}z_{(cb)}-\delta_{bc}z_{(ad)}+\delta_{bd}z_{(ca)}\right]\delta(x,y)\ne 0,\\
\{L_{a0}(x),{\cal S}_i(y)\}&=e\tilde{f}_{bi}\left[\beta_1\delta_{ba}e_0(z)+\beta_2(\delta_{ba}e_1(z)-z_{ba})\right.\nonumber\\
&\left.+\beta_3(\delta_{ba}e_2(z)+z_{bc}z_{ca}-zz_{ba})\right]\delta(x,y)\ne 0.
\end{align}
All these constraints, besides ${\cal S}_i$, have no analogs in the metric approach. They serve to manage the variables that are necessary for the tetrad approach but absent in the metric one. 
The special role of constraints  ${\cal S}_i$ in the metric approach is that the Lagrange multipliers  $u_i$ are determined from them (or with the help of them the Dirac brackets are defined), and these  $u_i$ are supposed to be in one-to-one correspondence with Hassan-Rosen new variables $n^i$ (i.e. $v^i$ in this work). 

Next, we exclude $u^i$ from the Hamiltonian with the help of Eq.~(\ref{eq:u_iv_i}). Also, we omit
 second class constraints specific to tetrads  (\ref{eq:Lminus}), (\ref{eq:L0}) from the Hamiltonian, then the Hamiltonian may be represented in a symmetric (concerning both metrics) form as follows
\be
\mathrm{H}=\int d^3x\left(  
N{\cal R}'+\bar N{\cal S}+N^i{\cal R}_i+\lambda^{+}_{ab}L^+_{ab}
 \right).
\ee
 Here the constraints standing at the lapse functions are the following
\be
{\cal R}'={\cal R}''+v^i{\cal S}_i,\qquad {\cal S}={\cal S}'+\bar{v}^i{\cal S}_i.\label{eq:RprimeS}
\ee
In the form preferring metric $f_{\mu\nu}$ used in our spacetime basis $(n^\alpha,e^\alpha_i)$ the Hamiltonian is as follows
\be
\mathrm{H}=\int d^3x\left(  
N{\cal R}+N^i{\cal R}_i+\lambda^{+}_{ab}L^+_{ab}
 \right).\label{eq:Hshort}
\ee
One may represent constraints (\ref{eq:RprimeS}) in other form\footnote{By putting second class constraints equal to zero we get  ${\cal R}={\cal R}'={\cal R}''$.}
\begin{alignat}{2}
{\cal R}'&={\cal H}+v^i\bar{\cal H}_i+\frac{2m^2}{\kappa}W',&\quad\mbox{where}\quad &W'=W+v^iV_i,\\
{\cal S}&=\bar{\cal H}+\bar{v}^i\bar{\cal H}_i+\frac{2m^2}{\kappa}V',&\quad\mbox{where}\quad &V'=V+\bar{v}^iV_i.
\end{alignat}
By straightforward calculation of Poisson brackets one can check that algebra (\ref{eq:begin}) -- (\ref{eq:SO}) written below is valid for
 the constraints that compose Hamiltonian (\ref{eq:Hshort}). 
In particular, functions ${\cal H}_k$ generate spatial coordinate transformations for the expressions constructed of metric $f_{\mu\nu}$,
\begin{align}
\{{\cal R}_k(x),{L}_{a0}(y)\}&\equiv
\{{\cal H}_k(x),{L}_{a0}(y)\}={L}_{a0}(x)\delta_{,k}(x,y)\approx 0,\label{eq:begin}\\
\{{\cal R}_k(x),{L}_{ab}(y)\}&\equiv
\{{\cal H}_k(x),{L}_{ab}(y)\}={L}_{ab}(x)\delta_{,k}(x,y)\approx 0,\\
\{{\cal R}_k(x),{\cal H}(y)\}&\equiv
\{{\cal H}_k(x),{\cal H}(y)\}={\cal H}(x)\delta_{,k}(x,y).
\end{align}
Functions $\bar{\cal H}_k$ do the same for the corresponding expressions constructed of $g_{\mu\nu}$,
\begin{align}
\{{\cal R}_k(x),\bar{L}_{ab}(y)\}&\equiv
\{\bar{\cal H}_k(x),\bar{L}_{ab}(y)\}=\bar{L}_{ab}(x)\delta_{,k}(x,y),\\
\{{\cal R}_k(x),\bar{\cal H}(y)\}&\equiv
\{\bar{\cal H}_k(x),\bar{\cal H}(y)\}=\bar{\cal H}(x)\delta_{,k}(x,y).
\end{align}
The Poisson brackets between ${\cal R}$ and ${\cal R}_i$ give the standard algebra of hypersurface deformations when the second class constraint ${\cal S}$ is taken into account
\begin{align}
\{{\cal R}(x),{\cal R}(y)\}&=\left(\eta^{ik}{\cal R}_k+uu^i{\cal S}\right)(x)\delta_{,i}(x,y)\nonumber\\
&-\left(\eta^{ik}{\cal R}_k+uu^i{\cal S}\right)(y)\delta_{,i}(y,x),\label{eq:RR}\\
\{{\cal R}_i(x),{\cal R}(y)\}&={\cal R}(x)\delta_{,i}(x,y)+u_{,i}{\cal S}\delta(x,y),\\
\{{\cal R}_i(x),{\cal R}_j(y)\}&={\cal R}_j(x)\delta_{,i}(x,y)-{\cal R}_i(y)\delta_{,j}(y,x).
\end{align}
Poisson brackets of the second class constraints ${\cal S}$, $\Omega$ provide the conditions for cancellation of the ghost degree of freedom
\begin{align}
\{{\cal S}(x),{\cal S}(y)\}&={\bar v}^i{\cal S}(x)\delta_{,i}(x,y)-{\bar v}^i{\cal S}(y)\delta_{,i}(y,x),\label{eq:SS}\\
\{{\cal R}(x),{\cal S}(y)\}&=(u^i+u\bar{v}^i){\cal S}(x)\delta_{,i}(x,y)+\left(u(\bar{v}^i{\cal S})_{,i}-\Omega\right)\delta(x,y),\label{eq:SR}\\
\{{\cal S}(x),\Omega(y)\}&\ne 0.\label{eq:SO}
\end{align}
Eqs. (\ref{eq:SS}) -- (\ref{eq:SO}) give the most important set of the second class constraint algebra. The primary constraint  ${\cal S}$  should commute with itself for appearance of the secondary constraint $\Omega$ from the compatibility condition $\dot{\cal S}=\{S(x), {\rm H}\}\approx 0$. The only one of the Poisson brackets  (\ref{eq:SS}) -- (\ref{eq:SO})  is nonzero on the constraints surface. It is necessary for the constraints ${\cal S}$ and $\Omega$ to form a pair of second class constraints. As $\Omega$ appears in Eq.~(\ref{eq:SR}) being multiplied on the $\delta$-function the lapse function  $N$ appears in the evolutionary equation that is necessary to preserve the constraint  ${\cal S}$ as a multiplier of constraints
\be
\dot{\cal S}=\{S(x), {\rm H}\}= N(x)\Omega(x)+\Bigl((N^i+Nv^i)(x){\cal S}(x)\Bigr)_{,i}\approx 0, 
\ee
and therefore stays arbitrary.

It is interesting to compare  Eqs.(\ref{eq:RR}) -- (\ref{eq:SR}) with  similar relations from the work~\cite{SolTch2}, derived without using any explicit expression of the dRGT potential, and based on the Dirac brackets: see Eqs.~(48) -- (51) of that work and also the non numbered equation preceding to Eq.~(55). The results are in full agreement. 

The Lagrange multipliers standing at the first class constraints, $N$, $N^i$, $\lambda^+_{ab}$, are arbitrary functions of time  $\tau$ and of spatial coordinates $x^i$. The variable $u$ can be determined from the following equation
\be
\dot\Omega=0\approx\int d^3x N\left(\{\Omega,{\cal R}'\}+u\{\Omega,{\cal S}\} \right),
\ee 
which is linear in  $u$.

The Hassan-Rosen transform takes in the tetrad formalism the following form

\be
u^i=v^i+u{\bar v}^i,
\ee
and arises simply as a solution for a Lagrange multiplier, this helps to avoid any need for the Dirac brackets.

\section{Conclusion}
In the tetrad Hamiltonian formalism of bigravity, we have  $n=21\times 2$  canonical variables, $({\bf e}_{ai},\pi^{i}_a)$, $(p_i,\Pi^i_0)$, $({\tilde f}_{ai},\Pi^i_a)$, $n_{fc}=7$ first class constraints, ${\cal R}$, ${\cal R}_i$, $L^+_{ab}$, and $n_{sc}=14$ second class constraints, ${\cal S}$, $\Omega$, $L^-_{ab}$, $G_{ab}$, $L_{a0}$, $S_i$. The calculation of the gravitational degrees of freedom gives us the well-known number
\be
n_{DOF}=\frac12(n-2n_{fc}-n_{sc})=7.
\ee 

The dynamics of bigravity is expressed through the algebra of its const\-raints. Comparing our results with the other calculations we see that this algebra does not depend on the approach and on the choice of variables, in particular this is true for the key pair of second class constraints ${\cal S}$, $\Omega$, modulo other second class constraints. These derivations do not require using  Dirac brackets, they are based on standard Poisson brackets, after their calculation the second class constraints are taken into account, especially $z_{ab}=z_{ba}$, or equivalently, $z_{ij}=z_{ji}$,  and 
\be
\frac{\delta \mathrm{H}}{\delta u^i}= {\cal S}_i\equiv {\bar {\cal H}}_i+\frac{2m^2}{\kappa}V_i=0.\label{eq:S_i}
\ee
Here we emphasize  the difference from the approach of works~\cite{HaRo,H-L}, as there the change of variable  $u^i=n^i+uD^i_{\ j}n^j$
is done before, and so instead, of equating   $\frac{\delta \mathrm{H}}{\delta u^i}$ to zero,  whereas  Eq.~(\ref{eq:S_i}) is argued there to be a consequence of the variation in a new variable $\frac{\delta \mathrm{H}}{\delta n^i}=0$.

As it was mentioned long ago in studying the Hamiltonian structure of the GR~\cite{Teit1973, HKT} the constraint algebra  gives  important information on the theory. Any reasonable modification of the GR preserving the general covariance should give the first class constraints satisfying the same algebra. The appearance of canonical variables in the coefficients of the algebra allows to get information on the Hamiltonian.

With the second class constraints coming into play  the algebra becomes more involved, but these constraints may be put to zero \emph{after} the calculation of Poisson brackets. 
For the first class quantities, here all of them are first class constraints, one may use the Poisson brackets and take into account the second class constraints after this calculation. Also it is possible to make calculations step by step when first one defines intermediary Dirac brackets made of the subset of the second class constraints. In a sense, a similar procedure is used here, as we preserve a couple of second class constraints ${\cal S}$, $\Omega$ in Eqs.~ (\ref{eq:RR}) -- (\ref{eq:SR}) and at the same time treat other second class constraints $L^-_{ab}$, $G_{ab}$, $L_{a0}$, $S_i$ as equal to zero. 
This method for derivation the algebra of the most essential for the bigravity constraints seems the most natural and simple among the proposed before. As it was shown for the first time in the work~\cite{HiRo}, the tetrad variables allow expressing the dRGT potential as a linear combination of all nonzero external products of the tetrad 1-forms.  It was also proved there that one needs the symmetry conditions
 for the equivalence between the metric and the tetrad approaches. Here the parameter $p_a=\tilde{v}_a $ introduced for  the nonsuited tetrad is treated as a function of canonical variables  and has nonzero Poisson brackets. This considerably simplifies the calculations.

The author thanks Yukawa Institute for Theoretical Physics at Kyoto University,
where this work was initiated during the workshop YITP-T-17-02 ``Gravity and Cosmology 2018''. It is also a pleasure to express his gratitude to the Referees of this work for their proposals  on improving the manuscript.

\newpage
\section*{Appendix A}\label{S:A}
The same problems were already considered in our previous work~\cite{SolTetrad}, but  published~\cite{SolTetrad, SolTetrad2} results were limited to the case of minimal potential ($\beta_1\ne 0$, $\beta_2=\beta_3=0$) as that method of Poisson brackets calculations was more involved. The notations and canonical variables applied here are to simplify the work. Appendix B is added  to compare our notations with the ones used by other authors.

The potential that couples two metrics in the tetrad formalism given the symmetry of tetrads conditions $z_{ab}=z_{ba}$ following from Eq.~(\ref{eq:u^izamena2}) is as follows 
\begin{align}
\tilde U&=uV+u^iV_i+W,\\
V&=e\left(\beta_0e_0(z)+\beta_1e_1(z)+\beta_2e_2(z)+\beta_3e_3(z) \right),\\
V_i&= -{\bf f}_{ia}C_{ab}p_b, \\
W&=e\left(\beta_0e_0(w)+\beta_1e_1(w)+\beta_2e_2(w)+\beta_3e_3(w) \right),
\end{align}
where $e=\det({\bf e}_{ai})$, $f=\det({\bf f}_{ai})$, $u=\frac{\bar N}{N}$, $u^i=\frac{\bar{N}^i-N^i}{N}$, $e_i$ are symmetric polynomials of   $(3\times 3)$-matrices $z_{ab}$, $w_{ab}$, $x_{ab}$ given below
\begin{align}
z_{ab}&=
{\cal P}_{ac}x_{cb}
\equiv 
\tilde{f}_{ia}{\bf e}^{ib}, 
\qquad {\cal P}_{ac}=\delta_{ac}+\frac{p_ap_c}{\varepsilon+1},\\
w_{ab}&={\cal P}^{-1}_{ac}x_{cb}
\equiv 
{\tilde f}^{ia}\eta_{ij}{\bf e}^{jb}, 
\qquad {\cal P}^{-1}_{ac}=\delta_{ac}-\frac{p_ap_c}{\varepsilon(\varepsilon+1)},\\
x_{cb}&=
{\bf f}_{ic}{\bf e}^{ib},
\qquad 
\tilde{f}_{ia}={\cal P}_{ac}{\bf f}_{ic},
\qquad {\tilde f}^{ia}={\cal P}^{-1}_{ac}{\bf f}^{ic},\\
C_{ab}&=
e\left[\beta_1\delta_{ba}e_0(x)+\beta_2(\delta_{ba}e_1(x)-x_{ba})\right.\\
&\left.+\beta_3(\delta_{ba}e_2(x)+x_{bc}x_{ca}-xx_{ba})
\right].
\end{align}
After substitution  of the expression found for the Lagrange multiplier $u^i=v^i+u{\bar v}^i$ which also follows from Eq.~(\ref{eq:u^izamena}) the potential simplifies and becomes the  following
\be
\tilde U=uV'+W',
\ee
where
\begin{align}
V'&=e\Bigl(\beta_1e_1(w)+\beta_2e_2(w)+\beta_3e_3(w) \Bigr)+\beta_0e,\\
W'&=\frac{e}{\varepsilon}\Bigl(\beta_1e_0(z)+\beta_2e_1(z)+\beta_3e_2(z) \Bigr)+\beta_4f.
\end{align}
This form of the potential is equivalent to the analogous formulas (2.23) -- 

By estimating Poisson brackets in Eq.~(\ref{eq:SR}) one obtains the secondary constraint  
\begin{align}
\Omega&=\frac{2m^2}{\kappa}\Biggl[ \frac{\partial{\cal H}}{\partial \Pi^i_a}\frac{\partial V'}{\partial\tilde{f_{ai}}}-\frac{\partial\bar{\cal H}}{\partial\pi^i_a}\frac{\partial W'}{\partial e_{ai}}-\nonumber\\
&-v^i\tilde{f}_{ak,i}\frac{\partial V'}{\partial\tilde{f}_{ak}}-\bar{v}^ie_{ak,i}\frac{\partial W'}{\partial e_{ak}}+\nonumber\\
&+v^i_{,k}\left(e_{ai}\frac{\partial }{\partial e_{ak}}-\delta^k_i\right)V'  
+\bar{v}^i\left(e_{ai}\frac{\partial W'}{\partial e_{ak}} \right)_{,k}
\Biggr].
\label{eq:Omega}
\end{align}
which forms a pair of second class constraints together with ${\cal S}$. This pair just excludes the ghost degree of freedom.

\section*{Appendix B}\label{S:MayDay}
 As each research group exploits a lot of special notations we hope it would be useful to add dictionaries for translations between them. We  include this Appendix to make  a reading of this paper easier for someone  familiar with the others.

\newpage
\renewcommand{\hoffset}{\hoffset=-50mm}
\begin{table}
\setlength{\extrarowheight}{5pt}
\begin{tabular}{|c||c|c|}
\hline 
{\bf variables}  & {\bf this work} & {\bf Alexandrov} \\
\hline
space-time coordinate 4-indices & $\alpha,\beta,\ldots,\mu,\nu$ & 
$\mu,\nu$ \\
\hline
spatial coordinate 3-indices  & $i,j,k...$ & $a,b,c$\\
\hline
internal 4-indices & $A,B,C\ldots$ & $I,J,K\ldots$ \\
\hline
internal 3-indices & $a,b,c\ldots$ & $i,j,k\ldots$\\
\hline 
1st space-time tetrad & $F^A_\mu$ & $e^K_+$\\
\hline
2nd space-time tetrad& $E^A_\mu$ & $e^K_-$  \\
\hline
internal 4-metric 
&$\eta_{AB}$ 
& $\eta_{IJ}$ \\
\hline
1st tetrad spatial components 
& $F^A_i=(p_i,\tilde{f}_{ai})$ 
& $e_{+a}^I=(E^j_{+,a}\chi_{+j},E^i_{+,a})$ \\
\hline
 2nd tetrad spatial components 
 & $E^A_i=(0,{\bf e}_{ai})$ 
 & $e_{-a}^I=(E^j_{-,a}\chi_{-j},E^i_{-,a})$ \\
\hline
tetrad normal components 
& $(F^A_\perp,E^A_\perp)$ 
& $(X^I_+,X^I_-)$ \\
\hline
Lorentz boost parameters 
& $(p_a,0)$ 
& $(\chi_{+,j},\chi_{-,j})$\\
\hline
1st induced spatial metric 
& $\eta_{ij}={\bf f}_{ia}{\bf f}_{aj}$ 
&not defined\\
\hline
2nd induced spatial metric 
& $\gamma_{ij}={\bf e}_{ia}{\bf e}_{aj}$ 
&not defined\\
\hline
hybrid spatial metric 
& $z_{ij}=\tilde{f}_{ia}{\bf e}_{aj}$
& ${\bf g}_{ab}=\eta_{IJ}e^I_{+,a}e^J_{-,b}$\\
\hline
and its determinant & $\det|z|=\varepsilon\det|{\bf f}|\det|{\bf e}|$
&$\det|{\bf g}|=\eta_{IJ}X^I_+X^J_-$\\
\hline
hybrid inverse spatial metric 
& $z^{ij}={\bf e}^{ia}\tilde{f}^{aj}$
& ${\bf g}^{ab}=E^a_{-,i}E^b_{+,j}\left(\delta^{ij}+\frac{\chi^i_+\chi^j_-}{1-\chi^k_+\chi_{-,k}}\right)$\\
\hline
\end{tabular}
\caption{Dictionary to translate variables between this work notations and notations of Ref.~\cite{Alex}. }

\end{table}

\renewcommand{\hoffset}{\hoffset=-50mm}
\begin{table}
\setlength{\extrarowheight}{5pt}
\begin{tabular}{|c||c|c|}
\hline 
{\bf variables}  & {\bf this work} & {\bf Kocic} \\
\hline 
1st spatial triad & ${\bf f}_{ai}$&$m=||m_{ai}||$ \\
\hline
2nd spatial triad &${\bf e}_{ai}$ & $e=||e_{ai}||$\\
\hline
2nd induced spatial metric 
& $\gamma_{ij}={\bf e}_{ia}{\bf e}_{aj}$ 
&$\gamma=e^{\rm T}\hat\delta e$\\
\hline
internal 4-metric 
&$\eta_{AB}$ 
&  \\
\hline
Lorentz boost parameters 
& $p_a$ 
&$p$ \\
\hline
& $\varepsilon=\sqrt{1+p_ap_a}=\frac{1}{\sqrt{1-v_av_a}}$&$\lambda$\\
\hline
Lorentz boost parameters 
& $v_a=\frac{p_a}{\varepsilon}$ 
& $v=\frac{p}{\lambda}$\\
\hline
Hassan-Rosen variable $n^i$&$v^i={\bf f}^{ia}v_a\equiv\tilde{f}^{ia}p_a$&$\tilde{n}=m^{-1} v$\\
\hline
Hassan-Rosen $D^i_{\ j}n^j$&$\bar{v}^i={\bf e}^{ia}v_a$&${n}$\\
\hline
&$P_{ab}=\delta_{ab}+\frac{p_ap_b}{\varepsilon+1}$ &$\hat\Lambda$ \\
\hline
&$P_{ab}p_b=\varepsilon p_a$&$\hat\Lambda p=\lambda p$\\
\hline
&$P^{-1}_{ab}=\delta_{ab}-\frac{p_ap_b}{\varepsilon(\varepsilon+1)}$ &$\hat\Lambda^{-1}$ \\
\hline
&$P^{-1}_{ab}p_b=\frac{1}{\varepsilon}p_a$&$\hat\Lambda^{-1}p=\frac{1}{\lambda} p$\\
\hline
&$\tilde{f}_{ai}=P_{ab}f_{bi}$&$\hat\Lambda m$\\
\hline
&$\tilde{f}^{ia}=P^{-1}_{ab}f^{bi}$&$m^{-1}\hat\Lambda^{-1} $ \\
\hline
nonsymmetric internal hybrid&$x_{ab}={\bf f}_{ai}{\bf e}^{ib}$&$me^{-1}$\\
\hline
symmetric internal hybrid&$z_{ab}=\tilde{f}_{ai}{\bf e}^{ib}$&$\hat\Lambda me^{-1}$\\
\hline
one more internal hybrid 
& 
$w_{ab}=\tilde{f}^{ia}\eta_{ij}{\bf e}^{jb}$
&
$\hat\Lambda^{-1}me^{-1}$\\
\hline
hybrid spatial metric 
& $z_{ij}=\tilde{f}_{ia}{\bf e}_{aj}$
&$m^{\rm T}\hat\Lambda e$ \\
\hline
its determinant & $\det|z|=\varepsilon\det|{\bf f}|\det|{\bf e}|$
&\\
\hline
hybrid inverse spatial metric 
& $z^{ij}={\bf e}^{ia}\tilde{f}^{aj}$
&$e^{-1}\hat\Lambda^{-1}(m^{-1})^{\rm T}$ \\
\hline
Hassan-Rosen matrix $D^i_{\ j}$&$z^{ik}\eta_{kj}$&$e^{-1}\hat\Lambda^{-1}m$\\
\hline
its symmetric polynomials & 
$\varepsilon^ke_k(w)$ & 
$e_k(D)$\\
\hline
1st space-time tetrad & $F^A_\mu$ & \\
\hline
2nd space-time tetrad& $E^A_\mu$ &   \\
\hline
1st induced spatial metric 
& $\eta_{ij}={\bf f}_{ia}{\bf f}_{aj}$ 
&$\phi=m^{\rm T}\hat\delta m$\\
\hline
1st tetrad spatial components 
& $F^A_i=(p_i,\tilde{f}_{ai})$ 
&  \\
\hline
 2nd tetrad spatial components 
 & $E^A_i=(0,{\bf e}_{ai})$ 
 &  \\
\hline
tetrad normal components 
& $(F^A_\perp,E^A_\perp)$ 
&  \\
\hline
\end{tabular}
\caption{Dictionary to translate variables between this work notations and notations of Ref.~\cite{Kocic}. }

\end{table}
\renewcommand{\hoffset}{\hoffset=-50mm}
\begin{table}
\setlength{\extrarowheight}{5pt}
\begin{tabular}{|c||c|c|}
\hline 
{\bf variables}  & {\bf this work} & {\bf Hassan -- Lundkvist} \\
\hline
& & \\
1st induced spatial metric 
& $\eta_{ij}={\bf f}_{ia}{\bf f}_{aj}$ 
&$\phi_{ij}$\\
\hline
& & \\
2nd induced spatial metric 
& $\gamma_{ij}={\bf e}_{ia}{\bf e}_{aj}$ 
&$\gamma_{ij}$\\
\hline
& & \\
1st metric lapse and shift  
& $(N,N^i)$
& $(L,L^i)$\\
\hline
& & \\
2nd metric lapse and shift & $(\bar{N},\bar{N}^i)$
&$(N,N^i)$\\
\hline
& & \\
difference of shifts 
& $\bar{N}^i-N^i=Nv^i+\bar{N}\bar{v}^i$
& $N^i-L^i=Ln^i+ND^i_{\ j}n^j$\\
\hline
& & \\
Hassan-Rosen variable & $v^i=\tilde{f}^{ia}p_a\equiv 
{\bf f}^{ia}v_a$
& $n^i$\\
\hline
& & \\
Hassan-Rosen matrix &$\varepsilon\tilde{f}^{ia}\eta_{jk}{\bf e}^{ka}$ & $D^i_{\ j}$\\
\hline
symmetric polynomials & $\varepsilon^ke_k(w)$ & $e_k(D)$ \\
\hline
Lorentz factor & $\varepsilon=\frac{1}{\sqrt{1-v^i\eta_{ij}v^j}}$& $x=1-n^i\phi_{ij}n^j$\\
\hline

&
$(vxv)=v_ax_{ab}v_b$
&
$n^i\phi_{ij}D^j_{\ k}n^k$
\\
\hline

&
$(vx^{\rm T}xv)=v_ax_{ba}x_{bc}v_c$
&
$D^i_{\ k}n^k\phi_{ij}D^j_{\ \ell}n^\ell$
\\
\hline
\end{tabular}
\caption{Dictionary for translation of notations  of this work  and  Ref.~\cite{H-L}. }

\end{table}

\renewcommand{\hoffset}{\hoffset=-50mm}
\begin{table}
\setlength{\extrarowheight}{5pt}
\begin{tabular}{|c||c|c|c|}
\hline 
{\bf constraints \&  multipliers} & {\bf this work} & {\bf Alexandrov}&{\bf Hassan-Lundkvist} \\
\hline
Hamiltonian constraint (1st class) 
& $N{\cal R}'$ 
&$N{\cal H}$ 
& $L\tilde{R}^0$\\
\hline
3-diff  generator (1st class)
& $N^i{\cal R}_i$
&   $N^a{\cal D}_a$
& $L_i\tilde{R}_i $\\
\hline
diagonal Gauss (1st class)& $\lambda^{+}L^+_{ab}$ & $n^{IJ}{\cal G}_{IJ}$&not defined \\
\hline
off-diagonal Gauss (2nd class) & $\lambda^-_{ab}L^-_{ab}$ & $\hat{n}^{IJ}\hat{\cal G}_{IJ}$&not defined\\
\hline
off-diagonal Gauss (2nd class)  & $\lambda^aL_{a0}$ &  $\hat{n}^{IJ}\hat{\cal G}_{IJ}$&not defined \\
\hline
2nd class 
&$u^i{\cal S}_i$ 
& $\hat{N}^a\hat{\cal D}_a$ &${\cal C}_k$\\
\hline
2nd class 
& $u{\cal S}$ 
& $\hat{N}\hat{\cal H}$ &$N{\cal C}$\\
\hline 
2nd class  (secondary)& $G_{ab}$ or $z_{ij}=z_{ji}$
& $S^{a}$&not defined\\
\hline
2nd class  (secondary)
 & $\Omega$ 
 & $\Psi$ &${\cal C}_2$\\
\hline
\end{tabular}
\caption{Dictionary to translate the 1st and 2nd class constraints between this work notations and notations of Refs.~\cite{Alex,H-L}. }
\end{table}


\begin{thebibliography}{**}



\bibitem{dRGT}
C.~de~Rham, G.~Gabadadze, and A.~J.~Tolley,
{\it Phys. Rev. Lett.} {\bf 106} 231101 (2011), 	arXiv:1011.1232.

\bibitem{dRGT2}
C.~de~Rham, G.~Gabadadze, and A.~J.~Tolley,
{\it Phys. Lett. B}
{\bf 711} 190-195 (2012), arXiv:1107.3820.

\bibitem{Reviews}
C.~de~Rham, 
  {\emph{Living Rev. Rel.} {\bf
  17} 7 (2014)}, 
  {arXiv:1401.4173}.


\bibitem{HaRo}
S.~F.~Hassan and R.~A.~Rosen, 
{\it Phys. Rev. Lett.} {\bf 108} 041101 (2012),	arXiv:1106.3344.

\bibitem{HaRo2}
	S.~F.~Hassan, R.~A.~Rosen, and A.~Schmidt-May,
	 	{\it JHEP} {\bf 1202} 026 (2012), arXiv:1109.3230.
	 	
\bibitem{HaRo3}	 	
	 	S.~F.~Hassan and R.~A.~Rosen,
{\it JHEP}
{\bf 1202} 126 (2012),	 	 	arXiv:1109.3515.

\bibitem{HaRo4}
S.~F.~Hassan and R. A. Rosen, 
{\it JHEP} {\bf 1204}  123 (2012),
	arXiv:1111.2070.
	
\bibitem{HiRo}
K.~Hinterbichler and R.~A.~Rosen,
{\it JHEP} {\bf 07} 047 (2012),
arXiv:1203.5783.

\bibitem{Dirac1950}
P.~A.~M.~Dirac,
{\it Can. J. Math.} {\bf 2} no. 2 129-148 (1950).

\bibitem{Dirac}
P.~A.~M.~Dirac. Lectures on quantum mechanics.\\
New York: Belfer Graduate School of Science, Yeshiva University,
1964.

\bibitem{ADM}
R.~Arnowitt, S.~Deser, and Ch.W.~Misner. The Dynamics of General Relativity. \\
In: Gravitation, an Introduction
to Current Research. ed. L.~Witten,  Wiley, New York (1963), arXiv:gr-qc/0405109. 



\bibitem{Krasnov}
S.~Alexandrov, K.~Krasnov, and S.~Speziale, {\it JHEP} {\bf 06} 068 (2013),
arXiv:1212.3614.


\bibitem{Alex}
S.~Alexandrov,
\textit{Gen.Rel.Grav.} {\bf 46} 1639 (2014), arXiv:1308.6586.


\bibitem{Kluson_tetrad}
J.~Kluson,
{Hamiltonian Formalism of Bimetric Gravity In Vierbein Formulation.} 
\textit{Eur. Phys.~J.} {\bf C 74} 2985 (2014), arXiv:1307.1974.


\bibitem{SolTch1}
V.~O.~Soloviev and M.~V.~Tchichikina,
\textit{Theor. Math. Phys.}, {\bf 176}:3 1163-1175 (2013), arXiv:1211.6530. 

DOI: https://doi.org/10.1007/s11232-013-0097-y

\bibitem{SolTch2}
V.~O.~Soloviev and M.~V.~Tchichikina,
\textit{Phys. Rev.}, {\bf D88}, 084026 (2013), arXiv:1302.5096. 

DOI: https://doi.org/10.1103/PhysRevD.88.084026

\bibitem{SolTetrad}
V.~O.~Soloviev,
\textit{Theor. Math. Phys.},  {\bf 182}:2, 294-307 (2015), arXiv:1410.0048. 

DOI: https://doi.org/10.1007/s11232-015-0263-5

\bibitem{SolTetrad2}
V.~O.~Soloviev,
Bigravity in tetrad Hamiltonian formalism and matter couplings.
 arXiv:1410.0048. 

\bibitem{SolCosmol}
V.~O.~Soloviev,
\textit{Phys. Part. Nuclei}, {\bf 48}:2, 287-308 (2017), arXiv:1505.00840. 

DOI: https://doi.org/10.1134/S1063779617020071



\bibitem{Kocic}
M.~Kocic.
{Geometric Mean of Bimetric Spacetimes.}
arXiv:1803.09752v1.

\bibitem{H-L}
S.~F.~Hassan and A.~Lundkvist,
{\it JHEP} {\bf 08} 182 (2018), 
arXiv:1802.07267v1.

\bibitem{Dirac1958}
P.~A.~M.~Dirac,
{\it Proc. Roy. Soc.} {\bf A246} 333-343 (1958).

\bibitem{Dirac1959-1}
P.~A.~M.~Dirac,
{\it Phys. Rev.} {\bf 114} 924-930 (1959).

\bibitem{Dirac1959-2}
P.~A.~M.~Dirac,
{\it Phys. Rev. Lett.} {\bf 2} 368-371 (1959).





\bibitem{ADM1959}
R.~Arnowitt, S.~Deser, and Ch.W.~Misner, 
{\it Phys.Rev. } {\bf  116}  1322-1330 (1959).

\bibitem{ADM1960}
R.~Arnowitt, S.~Deser, and Ch.W.~Misner,
{\it Phys.Rev. } {\bf  117}  1595-1602 (1960).



\bibitem{Kuchar1973}
 K.~Kucha\u{r},  
 Canonical Quantization of Gravity.
 In: Relativity, Astrophysics and Cosmology. Proceedings of the Summer School Held, 14 -- 26 August 1972, at the Banff Centre, Banff, Alberta, ed. W.~Israel, D.~Reidel Publishing Company, Dordrecht-Holland/ Boston-U.S.A., 1973.

\bibitem{Kuchar}
 K.~Kucha\u{r},  {\it J.~Math.~Phys.} 
 {\bf 17} 777-791 (1976). 
 
 \bibitem{Kuchar2}
 K.~Kucha\u{r},  {\it J.~Math.~Phys.} 
 {\bf 17}  792-800 (1976).
 
 \bibitem{Kuchar3}
 K.~Kucha\u{r},  {\it J.~Math.~Phys.} 
 {\bf 17}  801-820 (1976).
 
\bibitem{Kuchar4}
 K.~Kucha\u{r},  {\it J.~Math.~Phys.} 
{\bf 18} 1589-1597 (1977).

\bibitem{York}
J.~W.~York, Jr. Kinematics and Dynamics of General Relativity.\\
In: Sources of the Gravitational Radiation, edited by L.L. Smarr, 
Proceedings of the Battelle Seattle Workshop,
July 24 -- August 4, 1978.
Center for Astrophysics and Lyman Laboratory of Physics
Harvard University.


\bibitem{Sol1988}
V.O.~Soloviev,
{\it Sov. J. Part. Nucl.} {\bf 19} 482-497 (1988). 


\bibitem{Gourg}
E.~Gourgoulhou. $3+1$ Formalism and Bases of Numerical Relativity. gr-qc/0703035v1

\bibitem{Kocic2}
M.~Kocic,
{\it JHEP} {\bf 10} 219 (2019),
arXiv:1804.03659v1.


\bibitem{Hassan2014}
S.~F.~Hassan, M.~Kocic, and A.~Schmidt-May,
Absence of ghost in a new bimetric-matter coupling,
arXiv:1409.1909.







\bibitem{NRosen}
 N.~Rosen, 
{\it Phys. Rev.} 
{\bf 57} 147-150 (1940). 


\bibitem{NRosen2}
 N.~Rosen, 
{\it Phys. Rev.} 
{\bf 57} 150-153 (1940).


\bibitem{Salam} A.~Salam, and J.~Strathdee, {\it Phys. Rev.} {\bf 184} 1750-1759 (1969). 


\bibitem{Salam2} A.~Salam, and J.~Strathdee, {\it Phys. Rev.} {\bf 184}
1760-1768 (1969).

\bibitem{Salam3}
 C.~J.~Isham, A.~Salam, and J.~Strathdee, {\it Phys. Lett. B} {\bf 31} 300-302 (1970). 
 
 \bibitem{Salam4}
 C.~J.~Isham, A.~Salam, and J.~Strathdee,
 {\it Phys. Rev.} {\bf D3} 867-873 (1971).

\bibitem{WZ} B.~Zumino. Effective Lagrangians and broken symmetries.
\\
In: Brandeis Univ. Lectures on Elementary Particles and Quantum Field Theory (MIT Press Cambridge, Mass.), Vol. 2,  1970, 437.

\bibitem{Kogan}
T.~Damour and I.~Kogan,
{\it Phys.Rev. } {\bf D 66}  104024 (2002).





\bibitem{BD}
D.~G.~Boulware and S.~Deser, {\it Phys.Rev. } {\bf D6} 3368-3382 (1972). 



\bibitem{Comelli2012}
D.~Comelli, M.~Crisostomi, F.~Nesti, and L. Pilo,
{\it Phys. Rev. D} {\bf 86} 101502(R) (2012),
arXiv:1204.1027.

\bibitem{Comelli2013}
D.~Comelli, F.~Nesti, and L.~Pilo,
{\it Phys. Rev. D} {\bf 87} 124021 (2013),
 arXiv:1302.4447.

\bibitem{Comelli_MayDay}
D.~Comelli, F.~Nesti, and L.~Pilo,
{\it JHEP} {\bf 07} 161 (2013),
arXiv:1305.0236.

\bibitem{Leznov}
D.~Fairlie, A.~Leznov,
{\it J. Geom. Phys.} {\bf 16} 385-390 (1995), arXiv:hep-th/9403134.



\bibitem{Kluson}
    J.~Kluson,  
{\it JHEP} {\bf 1201} (2012) 13,
arXiv:1109.3052.

\bibitem{Kluson:2012ps}
J.~Kluson, 
{\it Phys. Rev.} {\bf D87} 084017 (2013),
arXiv:1211.6267.

\bibitem{Kluson:2013}
J.~Kluson, 
{\it IJMP}  {\bf A 28}, No. 28, 1350143 (2013),
arXiv:1301.3296.

\bibitem{Kluson:2013-2}
J.~Kluson,
\textit{Eur. Phys.~J.} {\bf C 73} 2553 (2013),
arXiv:1303.1652.






\bibitem{HamTetrads}
{S.~Deser, C.~J.~Isham,}  \textit{Phys. Rev.} {\bf D14} 2505-2510 (1976).

\bibitem{HamTetrads2}
{J.~E.~Nelson, C.~Teitelboim,}  \textit{Ann. of Phys.} {\bf 116} 86-104 (1978).

\bibitem{HamTetrads3}
{M.~Henneaux,} 
\textit{Gen. Rel. Grav.} {\bf 9} 1031-1045 (1978). 

\bibitem{Henn83}
{M.~Henneaux,}
\textit{Phys. Rev.} {\bf D27} 986-989 (1983).


\bibitem{Teit1973}
C.~Teitelboim, \textit{Annals Phys.} {\bf 79} 542-557 (1973).

\bibitem{HKT}
S.~Hojman, K.~Kucha\u{r}, and C.~Teitelboim, \textit{Annals Phys.} {\bf 96} 88-135 (1976).

























\end{thebibliography}
\end{document}